\documentclass[prl,aps,twocolumn,epsfig,superscriptaddress,showpacs]{revtex4}
\usepackage{bm}
\usepackage{amsfonts}
\usepackage[dvips]{graphicx}
\usepackage{mathrsfs}
\usepackage[intlimits]{amsmath}
\usepackage[colorlinks=false,dvipdfm]{hyperref}

\begin{document}
\title{Quantifying Nonlocality Based on Local Hidden Variable Models}
\author{Dong-Ling Deng}
 \affiliation{Theoretical Physics
Division, Chern Institute of Mathematics, Nankai University, Tianjin
300071, People's Republic of China}
\author{Jing-Ling Chen}
 \email{chenjl@nankai.edu.cn}
\affiliation{Theoretical Physics Division, Chern Institute of
Mathematics, Nankai University, Tianjin 300071, People's Republic of
China}

\author{Zi-Sui Zhou}
\affiliation{College of Mathematics, Nankai University, Tianjin
300071, People's Republic of China}

\date{\today}

\begin{abstract}
We introduce a fresh scheme based on the local hidden variable
models to quantify nonlocality for arbitrarily high-dimensional
quantum systems. Our scheme explores the minimal amount of white
noise that must be added to the system in order to make the system
local and realistic. Moreover, the scheme has a clear geometric
significance and is numerically computable due to powerful
computational and theoretical methods for the class of convex
optimization problems known as semidefinite programs.
\end{abstract}


\maketitle

In a celebrated work in $1964$, Bell showed surprisingly that
different measurements on two separated parts of entangled quantum
states would lead to stronger correlations, which are unexplainable
by any local hidden variable (LHV) theory or shared randomness
only~\cite{1964Bell}. This kind of stronger correlations is now well
known as \textit{nonlocality}.

In the last decade, nonlocality has attracted much interest in both
theoretical and experimental
works~\cite{1998Weihs-nonlocalExperiments} not only because of its
close relations to the foundations of quantum mechanics, but also
because of the vital role it plays in many quantum computation and
information processes~\cite{1991Ekert-Cryptograph,2007Linden}. %
%
%
%
Nevertheless, we are still far away from fully understanding
nonlocality. Many fundamental questions still remain open, one of
which is how to quantify nonlocality. Up to now, there are various
schemes to quantify entanglement~\cite{1998Wootters,1996Bennett},
while few works for nonlocality quantification
exist~\cite{2009Gisin-QauntifyNonL}.

A nature consideration is to use the violations of Bell inequalities
as the measure of nonlocality, based on which some nonlocality
distillation protocols are proposed recently in the framework of
generalized nonsignaling theories~\cite{2009Forster-Brunner}.
However, as these distillation protocols show exemplarily the
advantages of this measure, they also drastically reveal its severe
shortcomings: (i) For a specific system, there are many tight Bell
inequalities that are not equivalent to each other. Then which
inequality is the best one as the measure of nonlocality for this
quantum system? (ii) Some quantum states do not violate a certain
Bell inequality, but they do violate some other inequalities. What
is worse, we even do not know whether these states violate Bell
inequalities since we in principle need to find out all the Bell
inequalities for the system to answer this question. (iii) Finding
out all tight Bell inequalities for a quantum system is a NP hard
problem~\cite{1989Pitowski}, so it is not pragmatic to use Bell
inequality as nonlocality measure. Due to these disadvantages and
the rapid theoretical and experimental developments in this realm,
it becomes more and more pressing to introduce a more reasonable
measure for nonlocality.


In this Letter, based on the LHV models, we introduce a possible
scheme with a clear geometric significance to quantify nonlocality
for arbitrarily high-dimensional quantum systems. This scheme
explores the minimal amount of white noise that must be added to the
system so that the new system is local and realistic. In addition,
by taking advantage of a brilliant work by Terhal \textit{et
al.}~\cite{2003Terhal} and the established method for semidefinite
programs, we show explicitly how to estimate efficiently this
quantification by
numerical method. 

To start with, we should specify some notations and definitions. Let
$\mathscr{H}^{[d]}=\mathscr{H}^{[d_1]}_1
\otimes\cdots\otimes\mathscr{H}^{[d_n]}_n$ denote a $d$-dimensional
Hilbert space of $n$ particle system. Here $\mathscr{H}^{[d_k]}_k$,
($k=1,\cdots,n$), denotes the $d_k$-dimensional Hilbert space of the
$k$th subsystem and $d=d_1\times\cdots\times d_k$. Denoting the set
of all quantum states on $\mathscr{H}^{[d]}$ that admit LHV models
by $\mathscr{L}_{lhv}$, then the scheme to quantify nonlocality is
as follows: For a specific quantum state $\rho$ on the Hilbert space
$\mathscr{H}^{[d]}$, let
$\rho_{\lambda}=\lambda\frac{I_d}{d}+(1-\lambda)\rho$,
($0\leq\lambda\leq1$), then the definition of the nonlocality is:
\begin{eqnarray}
\mathcal {N}(\rho)=\min_{\rho_\lambda\in\mathscr{L}_{lhv}} \lambda.
\end{eqnarray}
Here $I_d$ is a $d$-by-$d$ identity matrix. Obviously, this
definition of nonlocality has the following properties:

(i) $\mathcal{N}(\rho)$ is invariant under local unitary
transformations, i.e., $\mathcal{N}(\rho)=\mathcal{N}(\mathcal{U}_1
\otimes\cdots\otimes\mathcal{U}_n\rho\mathcal{U}^{\dagger}_1\otimes\cdots\otimes\mathcal{U}^{\dagger}_n)$.

(ii) $\mathcal {N}(\rho)=0$ if the state $\rho$ is separable. In
fact, if $\rho$ admits a HLV model, then it is obvious  $\mathcal
{N}(\rho)=0$.


(iii) The $\mathcal {N}(\rho)$ exists for any quantum state $\rho$
since $\rho_\lambda$ is separable as long as $\lambda$ is large
enough~\cite{1998Zyczkowski}.

(iv) $\mathcal {N}(\rho)<1$. Actually, for any state $\rho$  on
$\mathscr{H}^{[d]}$, $\rho_\lambda$ is separable for $\lambda\geq
1-1/ \sqrt{(d^2-1)(d-1)}$~\cite{2008Deng}. Thus we have a more
accurate upper bound $\mathcal {N}(\rho)\leq1-1/
\sqrt{(d^2-1)(d-1)}$.

(v) $\mathcal {N}(\rho)>0$ if $\rho$ is an entangled pure state.
This is obvious since any pure entangled state violates certain Bell
inequality~\cite{1991Gisin}.

\begin{figure}
\includegraphics[width=50mm]{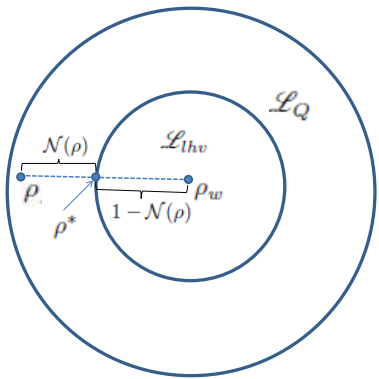}\\
 \caption{(Color online) A sketch to illustrate the geometric meaning
 of nonlocality. Here, the set of all quantum states is denoted by
 $\mathscr{L}_Q$;
 $\rho^*=\mathcal{N}(\rho)\frac{I_d}{d}+(1-\mathcal{N}(\rho))\rho$
 is the state with minimal $\lambda$ among all the states
 $\rho_{\lambda}$ that admit LHV models; $\rho_w=I_d/d$ denotes
  the white noise state (random state). If we normalize the distance
  between nonlocal state $\rho$ and the white noise state $\rho_w$, then the nonlocality
  $\mathcal{N}(\rho)$ indicates the distance between $\rho$ and $\rho^*$.}\label{fig1}
\end{figure}

Interestingly, this definition bears a geometric significance as
showed in the Fig.~\ref{fig1}. 
It is obvious that the nonlocality $\mathcal{N}(\rho)$ indicates the
distance between $\rho$ and $\rho^*$ if we normalize the distance
  between nonlocal state $\rho$ and the white noise state $\rho_w$. Moreover, from the experimental point of view,
$\mathcal{N}(\rho)$ also implies the minimal amount of white noise
that must be added to the system in order to hide the nonlocal
character of the state $\rho$. Its physical meaning is that it
provides a lower bound of how much noise is needed. In other words,
if the noise added to the system exceeds the bound, then the
nonlocal correlations of this system will disappear.

Now, we have the scheme to quantify nonlocality. However, from the
definition of $\mathcal{N}(\rho)$, it is very difficult to compute
$\mathcal{N}(\rho)$ analytically since the analytical construction
of LHV models for some entangled states is itself extremely
difficult~\cite{1989Werner}. Fortunately, Terhal \textit{et al}.
have introduced a simple and efficient algorithmic approach for
LHV-model construction for quantum states~\cite{2003Terhal}. Their
approach is based on the construction of a symmetric quasiextension
of the quantum state.
They showed explicitly that a symmetric
quasiextension might lead to a LHV model for the specific state,
depending only on the number of local measurement settings for each
observer. In the following, we shall show how to estimate
numerically $\mathcal{N}(\rho)$ by taking advantage of this method.
We will focus on the two-qubit and two-qutrit systems. The
generalization to multipartite higher-dimensional systems is
straightforward.

For convenience and completeness, we first recapitulate the
 main results of Terhal \textit{et al}.~\cite{2003Terhal}. Consider
the following Bell-type scenario: Two distant parties, Alice and
Bob, has a set of local measurements. Denote the number of
measurements for Alice as $\mathcal {M}_a$ and let each measurement
has $\mathcal {O}_a$ outcomes. Similarly, we denote the number of
measurements for Bob as $\mathcal {M}_b$ and let each measurement
has $\mathcal {O}_b$ outcomes. Moreover, let's define
\begin{eqnarray}
\mathcal {S}(\rho)=\frac{1}{\mathcal {M}!}\sum_{\Lambda}\Lambda\rho
\Lambda^{\dagger},
\end{eqnarray}
where $\Lambda$: $\mathscr{H}^{\otimes \mathcal
{M}}\rightarrow\mathscr{H}^{\otimes \mathcal {M}}$ is a permutation
of spaces $\mathscr{H}$ in $\mathscr{H}^{\otimes \mathcal {M}}$. We
say that $\rho$ on $\mathscr{H}_A\otimes\mathscr{H}_B$ has a
$(\mathcal {M}_a,\mathcal {M}_b)$-symmetric quasiextension when
there exists a multipartite entanglement witness $\mathcal
{W}_{\rho}$ on $\mathscr{H}_A^{\otimes \mathcal
{M}_a}\otimes\mathscr{H}_B^{\otimes \mathcal {M}_b}$ such that
$\texttt{Tr}_{\mathscr{H}_A^{\otimes \mathcal
{M}_a-1},\mathscr{H}_B^{\otimes \mathcal
{M}_b-1}}\mathcal{W}_{\rho}=\rho$ and
$\mathcal{W}_{\rho}=\mathcal{S}_A\otimes\mathcal{S}_B(\mathcal{W}_{\rho})$.
Terhal \textit{et al}. showed that if $\rho$ has a $(\mathcal
{M}_a,\mathcal {M}_b)$-symmetric quasiextension, then $\rho$
definitely admits a LHV model when Alice and Bob have
$\mathcal{M}_a$ and $\mathcal{M}_b$ arbitrary
measurements~\cite{2003Terhal}.

Now, the task of constructing LHV models is transformed to a new
task of finding symmetric quasi-extension for a specific state.
According to Ref.~\cite{2003Terhal}, the new task can be stated as a
particular case of the class of convex optimization known as
\textit{semidefinite programs} (SDP)~\cite{2003Terhal}, which
correspond to the optimization of a linear function subject to a
linear matrix inequality. Vandenberghe and Boyd
~\cite{1996Vandenberghe} write the typical SDP as:
\begin{eqnarray}
&\texttt{minimize}&\quad \mathbf{c}^T\mathbf{x},\nonumber\\
&\texttt{subject to}& \quad F(\mathbf{x})\geq 0,\nonumber
\end{eqnarray}
where $\mathbf{c}$ is a given vector of length $\nu$, and
$F(\mathbf{x})=F_0+\sum_ix_iF_i$; $F_i$ ($i=1,\cdots, \nu$) are some
fixed $\mu$-by-$\mu$ Hermitian matrices. The vector $\mathbf{x}$,
also of length $\nu$, is the variable over which the minimization is
performed. The SDP is said to be strictly feasible if there exists a
vector $\mathbf{x}$ such that $F(\mathbf{x})>0$ is satisfied. The
dual problem corresponding to a SDP, also a SDP, reads:
\begin{eqnarray}
&\texttt{maximize}& \quad -\texttt{Tr}[F_0Z],\nonumber\\
&\texttt{subject to}& \quad \texttt{Tr}[F_iZ]=c_i,\quad
Z\geq0,\nonumber
\end{eqnarray}
where the Hermitian matrix $Z$ is the variable over which the
maximization is performed. Similarly, if there exists a matrix $Z>0$
satisfying the trace constraints, the dual SDP is also said to be
strictly feasible. A very important relation between the primal and
dual optimizations was shown by Vandenberghe and Boyd
~\cite{1996Vandenberghe}: \textit{If both the primal and dual forms
of a SDP are strictly feasible, their optima are equal and achieved
by some feasible pair} ($\mathbf{x}_{opt},Z_{opt}$). The
constructions of symmetric extensions correspond to the dual form of
a SDP~\cite{2003Terhal} with $F_i=\mathcal
{S}_A\otimes\mathcal{S}_B(\lambda_i\otimes I)$,
$c_i=r_i=\texttt{Tr}[\lambda_i\rho]$ and $F_0=I_{d_A}\otimes
I_{d_B}\otimes I$. Here $\{\lambda_i\}$ is the basis for the space
of Hermitian matrices that operate on
$\mathscr{H}_A\otimes\mathscr{H}_B$. This basis is orthogonal in the
trace inner product $\texttt{Tr}[\lambda_i\lambda_j]=\delta_{ij}$
and $\lambda_0=I_{d_A}\otimes I_{d_B}/\sqrt{d_Ad_B}$. The
corresponding $\nu=(d_Ad_B)^2-1$ and $\mu=d_A^{\mathcal
{M}_a}d_B^{\mathcal {M}_b}$.

\begin{figure}
\includegraphics[width=94mm]{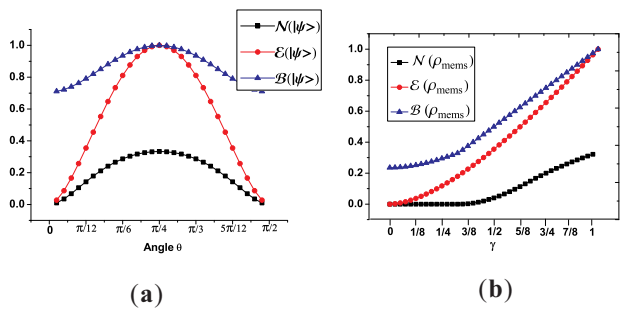}\\
 \caption{(Color online) Numerical results of nonlocality, Bell-CHSH inequality
 violation and entanglement of formation for: (a) the two-qubit
 pure state $|\psi\rangle=\sin\theta|00\rangle+\cos\theta|11\rangle$; (b)the maximally entangled mixed states
 $\rho_{mems}$.
 The nonlocality $\mathcal{N}(\rho)$ is estimated by
$\mathcal{N}^{[2,2]}(\rho)$ and the Bell-CHSH inequality violation
is normalized by dividing a factor of $2\sqrt{2}$.}\label{fig2}
\end{figure}

%
%

Based on the discussion above, in order to calculate
$\mathcal{N}[\rho]$, we only need to find the minimal $\lambda$ so
that $\rho_\lambda$ admits LHV model. Nevertheless, the above
numerical construction of LHV model depends on the number of local
measurement settings for each observer. Consequently, it might be
more convenient to introduce a transitional definition of
nonlocality also based on the number of local measurement settings
for each observer:
\begin{eqnarray}
\mathcal{N}^{[\mathcal{M}_a,\mathcal{M}_b]}(\rho)
=\min_{\rho_\lambda\in
\mathscr{L}^{[\mathcal{M}_a,\mathcal{M}_b]}_{lhv}}\quad\lambda,
\end{eqnarray}
where $\mathscr{L}^{[\mathcal{M}_a,\mathcal{M}_b]}_{lhv}$ is the set
of all the states that admit LHV models for
$(\mathcal{M}_a,\mathcal{M}_b)$-settings. We term
$\mathcal{N}^{[\mathcal{M}_a,\mathcal{M}_b]}(\rho)$ the
$(\mathcal{M}_a,\mathcal{M}_b)$-nonlocality of state $\rho$.
Mathematically, one has
\begin{eqnarray}\label{relationN-N}
\mathcal{N}(\rho)=\lim_{\mathcal{M}_a\rightarrow
\infty,\mathcal{M}_b\rightarrow
\infty}\mathcal{N}^{[\mathcal{M}_a,\mathcal{M}_b]}(\rho).
\end{eqnarray}
From the relationship stated in Eq. (\ref{relationN-N}), we can
always use $\mathcal{N}^{[\mathcal{M}_a,\mathcal{M}_b]}(\rho)$ to
estimate $\mathcal{N}(\rho)$ as long as $\mathcal{M}_a$ and
$\mathcal{M}_b$ are big enough. Actually, numerical results show
that for many states,
$\mathcal{N}^{[\mathcal{M}_a,\mathcal{M}_b]}(\rho)$ vary very
slightly with different $(\mathcal{M}_a,\mathcal{M}_b)$. So, we only
need to compute $\mathcal{N}^{[\mathcal{M}_a,\mathcal{M}_b]}(\rho)$
with small $\mathcal{M}_a$ and $\mathcal{M}_b$. Consequently, the
scheme to quantify nonlocality introduced above can be carried out
numerically due to the efficient method to construct LHV models with
$(\mathcal{M}_a,\mathcal{M}_b)$-settings.

To illustrate explicitly how the method works, we give some examples
here. We have implemented the mentioned corresponding semidefinite
program using SeDuMi~\cite{SeDuMi-Resorce}. The first example is
about a set of two-qubit states:
$|\psi\rangle=\sin\theta|00\rangle+\cos\theta|11\rangle$. The
nonlocality $\mathcal{N}(|\psi\rangle)$ with different $\theta$ is
showed in Fig.~\ref{fig2} (a). Note that here we use
$\mathcal{N}^{[2,2]}(|\psi\rangle)$ to estimate
$\mathcal{N}(|\psi\rangle)$. Also drawn are the entanglement of
formation (EOF)~\cite{1998Wootters} and Bell-CHSH~\cite{1969Clauser}
inequality violation of this set of states. From the figure, the
maximal nonlocality occurs at $\theta=\pi/4$, i.e., the maximally
entangled state
$|\Psi\rangle=\frac{1}{\sqrt{2}}(|00\rangle+|11\rangle)$, and the
maximal nonlocality $\mathcal{N}(|\Psi\rangle)$ equals to $1/3$. For
two-qubit pure state, the nonlocality  increases monotonically as
the EOF goes from $0$ to $1$. Interestingly, based on the fact that
$\mathcal{N}(|\Psi\rangle)=1/3$, we suspect that there might exist
some two-qubit Bell inequalities with an improved visibility $2/3$,
which is stronger than the CHSH inequality. However, we have not
found such an inequality and the question proposed by Gisin is still
open~\cite{2007Gisin}.

%

Another example involves the so called \textit{maximally entangled
mixed states} (MEMS), which have the maximum amount of entanglement
for a given linear entropy~\cite{2001Munro-MEMS}:
\begin{eqnarray}
\rho_{mems}=\left(\begin{matrix}
f(\gamma)&0&0&\gamma/2\\
0&1-2f(\gamma)&0&0\\
0&0&0&0\\
\gamma/2&0&0&f(\gamma)
\end{matrix}\right),
\end{eqnarray}
where $f(\gamma)=\left\{\begin{array}{ll} \gamma/2,&\quad
\texttt{for}\quad\gamma\geq2/3\\1/3,&\quad\texttt{for}
\quad\gamma<1/3\end{array}\right.$. 
The nonlocality $\mathcal{N}(\rho_{mems})$, entanglement of
formation $\mathcal{E}(\rho_{mems})$ and Bell-CHSH violation
$\mathcal{B}(\rho_{mems})$ for this set of states with different
$\gamma$ are plotted in Fig.~\ref{fig2} (b). 
From the figure, the difference between our scheme and the one using
Bell inequality violation to quantify nonlocality is obvious:
according to our quantification, the nonlocality of $\rho_{mems}$
arises when $\gamma>0.35$, while the Bell inequality approach fails
to quantify the nonlocality for
$0.35\leq\gamma\leq\frac{1}{\sqrt{2}}$ since the state does not
violate the CHSH inequality for this region. Moreover, it is
worthwhile to note that $\mathcal{E}(\rho_{mems})>0$ and
$\mathcal{N}(\rho_{mems})=0$ for the region $0<\gamma<0.35$,
indicating that for this region, the state $\rho_{mems}$ is
entangled, but it admits LHV model. This provides us a fresh
evidence from another aspect that entanglement and nonlocality are
different.

\begin{figure}
\includegraphics[width=84mm]{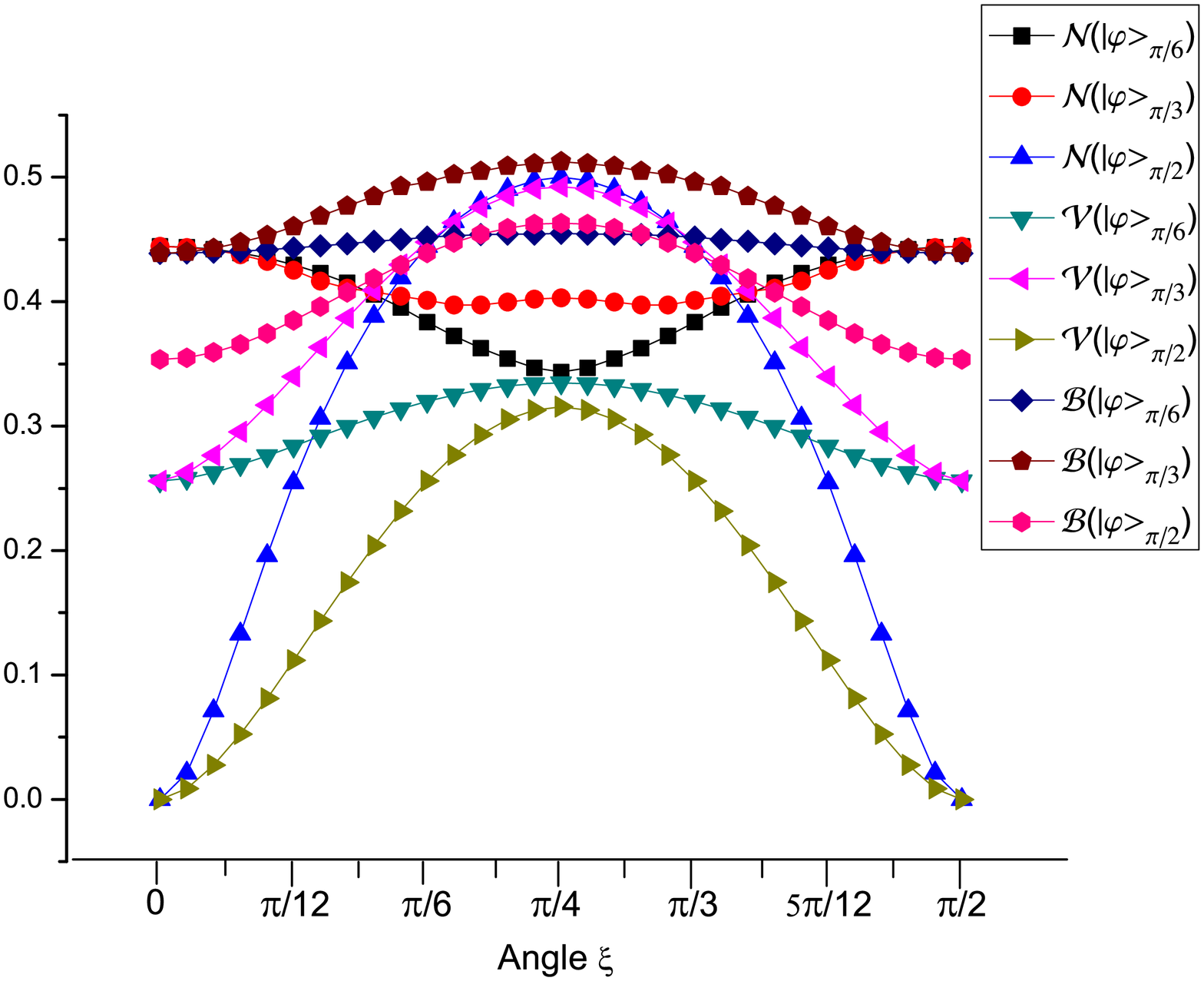}\\
 \caption{(Color online) Numerical results of nonlocality, Bell-CGLMP inequality
 violation and von Neumann entropy  for the two-qutrit
 state $|\phi\rangle=\sin\xi\sin\beta|00\rangle+\sin\xi\cos\beta|11\rangle+\cos\xi|22\rangle$.
 The nonlocality $\mathcal{N}(|\phi\rangle)$ is estimated by
$\mathcal{N}^{[2,2]}(|\phi\rangle)$. The Bell-CGLMP inequality
violation and the von Neumann entropy are normalized by dividing
$4\sqrt{2}$ and $2\ln3$, respectively. In this figure, we have only
plotted the curves with $\xi=\pi/6$, $\xi=\pi/3$ and $\xi=\pi/2$.
}\label{fig3}
\end{figure}

The third example concerns the two-qutrit system. Let us first
briefly introduce the Collins-Gisin- Linden-Massar-Popescu (CGLMP)
inequality~\cite{2002CGLMPBI} and the von Neumann entropy. For the
two-qutrit system, the CGLMP inequality reduces to:
\begin{eqnarray}\label{CGLMP-Qrtrits}
\mathcal{I}_3&=&[P(A_1\doteq B_1)+P(B_1\doteq A_2+1)+P(A_2\doteq
B_2)\nonumber\\
&+&P(B_2\doteq A_1)] -[P(A_1\doteq B_1-1)+P(B_1\doteq
A_2)\nonumber\\
&+&P(A_2\doteq B_2-1)+P(B_2\doteq A_1-1)]\leq2.
\end{eqnarray}
Here we use the symbol $\doteq$ to denote equality modulus $3$ and
denote the joint probability $P(A_i\doteq B_j+m)$ ($i, j=1, 2$) that
the measurements $A_i$ and $B_j$ have outcomes that differ, modulo
three, by $m$: $P(A_i\doteq B_j+m)=\sum_{a=0}^{2}P(A_i=a, B_j\doteq
a+m)$. 
%
%
Another related concept, which aptly measures the entanglement of
pure two-qutrit states, is the von Neumann entropy. For a pure state
$\rho_{AB}$ of two subsystems $A$ and $B$, the von Neumann entropy
of the reduced density operators is given by:
\begin{eqnarray}
\mathscr{V}(\rho_A)=\mathscr{V}(\rho_B)=-\texttt{Tr}[\rho_A\ln\rho_A]
=-\texttt{Tr}[\rho_B\ln\rho_B],
\end{eqnarray}
where $\rho_A=\texttt{Tr}_B[\rho_{AB}]$ and
$\rho_B=\texttt{Tr}_A[\rho_{AB}]$ are the reduced density matrices.
The state considered here is the generalized GHZ state of two-qutrit
system:
$|\phi\rangle=\sin\xi\sin\beta|00\rangle+\sin\xi\cos\beta|11\rangle+\cos\xi|22\rangle$.
As for the two-qutrit case, we also draw the nonlocality, von
Neumann entropy and the Bell-CGLMP inequality violation with
different $\xi$ and $\beta$ in Fig.~\ref{fig3}. Note that the
nonlocality $\mathcal{N}(|\phi\rangle)$ is estimated by
$\mathcal{N}^{[2,2]}(|\phi\rangle)$. For the convenience of drawing,
the Bell-CGLMP inequality violation and the von Neumann entropy of
the reduced density operators are divided by $4\sqrt{2}$ and
$2\ln3$, respectively. In other words,
$\mathcal{B}(|\phi\rangle)=\mathcal{I}_3^{Q}(|\phi\rangle)/4\sqrt{2}$
and $\mathcal{V}(|\phi\rangle)=\mathscr{V}(|\phi\rangle)/(2\ln3)$.
From the figure, $\mathcal{N}(|\phi\rangle)$,
$\mathcal{B}(|\phi\rangle)$ and $\mathcal{V}(|\phi\rangle)$ behave
quite differently as $\xi$ and $\beta$ vary. In this case, the
maximal nonlocality $\mathcal{N}(|\phi\rangle)$ does not occur at
the maximal entangled state. Actually, for the maximal two-qutrit
entangled state
$|\Phi\rangle=\frac{1}{\sqrt{3}}(|00\rangle+|11\rangle+|22\rangle)$,
the nonlocality is $\mathcal{N}(|\Phi\rangle)=\frac{3}{8}$, which is
less than $1/2$, the nonlocality 
of the state
$|\phi\rangle_{\xi=\pi/2,\beta=\pi/4}=\frac{1}{\sqrt{2}}
(|00\rangle+|11\rangle)$.

It is worthwhile to point out that for all the three examples, the
nonlocality $\mathcal{N}(\rho)$ is estimated by
$\mathcal{N}^{[2,2]}(\rho)$. Of course, one can use
$\mathcal{N}^{[\mathcal{M}_a,\mathcal{M}_b]}(\rho)$
($\mathcal{M}_a,\mathcal{M}_b>2$) as the approximation of
$\mathcal{N}(\rho)$ and the bigger $\mathcal{M}_a$ and
$\mathcal{M}_b$, the better the approximation. However, as
$\mathcal{M}_a$ and $\mathcal{M}_b$ become bigger, it might need
more time and EMS memory to run the SeDuMi program on the computer.
Actually, our numerical results show that for the states under
discussion, the variations of
$\mathcal{N}^{[\mathcal{M}_a,\mathcal{M}_b]}(\rho)$ with different
$\mathcal{M}_a$ and $\mathcal{M}_b$ are very small. Thus, we believe
$\mathcal{N}^{[2,2]}(\rho)$ is sufficient for a rough estimation of
$\mathcal{N}(\rho)$.

In summary, we introduced a reasonable scheme with a clear geometric
significance to quantify nonlocality based on the LHV models. The
scheme is numerically computable due to the systematic approach of
constructing LHV models for quantum states and the powerful
numerical methods for SDP problems. To illustrate how the method
works, we have provided some examples concerning two-qubit and
two-qutrit systems. Our approach does not need any Bell inequality,
thus it naturally circumvent the previously mentioned disadvantages
of using Bell inequality violation as nonlocality measure. This
scheme sheds a new light on the quantitative understanding of
quantum nonlocality. Interestingly, we have found that it may be
very useful in quantum phase transitions~\cite{2009Deng-QPT}. It
would also be interesting and significant to explore its possible
applications in quantum information and computation science, which
we shall investigate subsequently.

D. L. Deng is grateful for the hospitality of Kavli Institute for
Theoretical Physics China. This work was supported in part by NSF of
China (Grant No. 10975075), Program for New Century Excellent
Talents in University, the Project-sponsored by SRF for ROCS, SEM,
and the Project of Knowledge Innovation Program (PKIP) of Chinese
Academy of Sciences, Grant No. KJCX2.YW.W10.


\begin{thebibliography}{99}
\bibitem{1964Bell}
J. S. Bell, Physics (Long Island City, N. Y.) \textbf{1}, 195
(1964).

\bibitem{1998Weihs-nonlocalExperiments}
G. Weihs \emph{et al.}, Phys. Rev. Lett. \textbf{81}, 5039 (1998);
W. Tittel \emph{et al.}, Phys. Rev. Lett. \textbf{81}, 3563 (1998);
D. N. Matsukevich \emph{et al.}, Phys. Rev. Lett. \textbf{100},
150404 (2008).

\bibitem{1991Ekert-Cryptograph}
A. K. Ekert, Phys. Rev. Lett. \textbf{67}, 661 (1991); J. Barrett,
L. Hardy, and A. Kent, Phys. Rev. Lett. \textbf{95}, 010503 (2005);
A. Acin \emph{et al.}, Phys. Rev. Lett. \textbf{98}, 230501 (2007).

\bibitem{2007Linden}
N. Linden, S. Popescu, A. J. Short, and A. Winter, Phys. Rev. Lett.
\textbf{99}, 180502 (2007).



\bibitem{1996Bennett}
C. H. Bennett \emph{et al.}, Phys. Rev. A \textbf{54}, 3824 (1996);
V. Vedral, M. B. Plenio, M. A. Rippin, and P. L. Knight, Phys. Rev.
Lett. \textbf{78}, 2275 (1997).
\bibitem{1998Wootters}
 W. K. Wootters, Phys. Rev. Lett.
\textbf{80}, 2245 (1998).

\bibitem{2009Gisin-QauntifyNonL}
J. D. Bancal, C. Branciard, N. Gisin, and S. Pironio, Phys. Rev.
Lett. \textbf{103}, 090503 (2009).

\bibitem{2009Forster-Brunner}
M. Forster, S. Winkler, and S. Wolf, Phys. Rev. Lett. \textbf{102},
120401 (2009); N. Brunner and P. Skrzypczyk, Phys. Rev. Lett.
\textbf{102}, 160403 (2009).

\bibitem{1989Pitowski}
I. Pitowski, Quantum Probability, Quantum Logic (Springer Verlag,
Heidelberg, 1989).

\bibitem{2003Terhal}
B. M. Terhal, A. C. Doherty, and D. Schwab, Phys. Rev. Lett.
\textbf{90}, 157903 (2003).

\bibitem{1998Zyczkowski}
K. \.{Z}yczkowski, P. Horodecki, A. Sanpera, and M. Lewenstein,
Phys. Rev. A \textbf{58}, 883 (1998).

\bibitem{2008Deng}
D. L. Deng and J. L. Chen, arXiv:0810.2020v1

\bibitem{1991Gisin}
N. Gisin, Phys. Lett. A \textbf{154}, 201 (1991); N. Gisin and
Peres, Phys. Lett. A \textbf{162}, 15 (1992).

\bibitem{1989Werner}
R. F. Werner, Phys. Rev. A \textbf{40}, 4277 (1989).

\bibitem{1996Vandenberghe}
L. Vandenberghe and S. Boyd, SIAM Rev. \textbf{38}, 49 (1996).

\bibitem{SeDuMi-Resorce}
SeDuMi version 1.21 2009, available from
http://sedumi.ie.lehigh.edu.

\bibitem{1969Clauser}
J. Clauser, M. Horne, A. Shimony, R. Holt, Phys. Rev. Lett.
\textbf{23}, 880 (1969).

\bibitem{2007Gisin}
N. Gisin, arXiv:quant-ph/0702021v2.

\bibitem{2001Munro-MEMS}
W. J. Munro, D. F. V. James, A. G. White, and P. G. Kwiat, Phys.
Rev. A \textbf{64} 030302(R) (2001).

\bibitem{2002CGLMPBI}
D. Collins, N. Gisin, N. Linden, S. Massar, and S. Popescu, Phys.
Rev. Lett. \textbf{88}, 040404 (2002);

\bibitem{2009Deng-QPT}
D. L. Deng and J. L. Chen, in preparation.

%
%


%



%
%
%
%
%
%
%
%
%
%
%
%
























%
%
%
%
%
%
%
%
%
%
%
%
%
%
%
%
%
%
%
%
%
%
%
%
%
%
%
%
%
%
%
%
%

%
%
%
%


\end{thebibliography}
\end{document}